\documentclass[9pt,twocolumn,conference]{IEEEtran}
\usepackage{times}
\usepackage{amsbsy}
\usepackage{latexsym}
\usepackage{amsmath}
\usepackage[ansinew]{inputenc}
\usepackage[dvips ]{graphicx}
\input epsf
\usepackage{epic,eepic}
\usepackage{url}

\IEEEoverridecommandlockouts

\title{\huge Joint Iterative Power Adjustment and Interference Suppression
Algorithms for Cooperative DS-CDMA Networks \vspace{-0.5em}}
\author{Rodrigo C. de Lamare and Sheng Li \\ Communications Research Group \\ Department of Electronics,
    University of York, York Y010 5DD, United Kingdom \\
    Emails: \protect\url{rcdl500@ohm.york.ac.uk, sl546@york.ac.uk}
\thanks{\footnotesize The work of the authors was
supported by the University of York, York Y010 5DD, United
Kingdom.\vspace{-2.05em }}}

\begin{document}
\maketitle \thispagestyle{empty}

\begin{abstract}
This work presents joint iterative power allocation and
interference suppression algorithms for DS-CDMA networks which
employ multiple relays and the amplify and forward cooperation
strategy. We propose a joint constrained optimization framework
that considers the allocation of power levels across the relays
subject to individual and global power constraints and the design
of linear receivers for interference suppression. We derive
constrained minimum mean-squared error (MMSE) expressions for the
parameter vectors that determine the optimal power levels across
the relays and the parameters of the linear receivers. In order to
solve the proposed optimization problems efficiently, we develop
recursive least squares (RLS) algorithms for adaptive joint
iterative power allocation, and receiver and channel parameter
estimation. Simulation results show that the proposed algorithms
obtain significant gains in performance and capacity over existing
schemes. \vspace{+0.05em}
\end{abstract}


\section{Introduction}
Multiple collocated antennas enable the exploitation of the
spatial diversity in wireless channels, mitigating the effects of
fading and enhancing the performance of wireless communications
systems. Due to size and cost it is often impractical to equip
mobile terminals with multiple antennas. However, spatial
diversity gains can be obtained when terminals with single
antennas establish a distributed antenna array through cooperation
\cite{sendonaris}-\cite{laneman04}. In a cooperative transmission
system, terminals or users relay signals to each other in order to
propagate redundant copies of the same signals to the destination
user or terminal. To this end, the designer must employ a
cooperation strategy such as amplify-and-forward (AF)
\cite{laneman04}, decode-and-forward (DF) \cite{laneman04,huang}
and compress-and-forward (CF) \cite{kramer}.

Recent contributions in the area of cooperative and multihop
communications have considered the problem of resource allocation
\cite{luo,long}. However, there is very little work on resource
allocation strategies in multiuser spread spectrum systems. In
particular, prior work on cooperative multiuser spread spectrum
DS-CDMA systems in interference channels has not received much
attention and has focused on problems such as the impact of
multiple access interference (MAI) and intersymbol interference
(ISI), the problem of partner selection \cite{huang,venturino} and
the analysis of the bit error rate (BER) and outage performances
\cite{vardhe}. The problem of resource allocation and interference
mitigation in cooperative multiuser spread spectrum systems has
not been jointly dealt with so far. This problem is of paramount
importance in cooperative wireless ad-hoc and sensor networks
\cite{souryal,fischione} that utilize spread spectrum systems.

In this work, spread spectrum systems which employ multiple relays
and the AF cooperation strategy are considered. The problem of
joint resource allocation and interference mitigation in multiuser
DS-CDMA with a general number of relays is addressed. In order to
facilitate the receiver design, we adopt linear multiuser
receivers \cite{verdu} which only require a training sequence and
the timing. More sophisticated receiver techniques are also
possible \cite{verdu,delamaretc}. A joint constrained optimization
framework that deals with the allocation of power levels among the
relays subject to individual and global power constraints and the
design of linear receivers is proposed. MMSE expressions that
jointly determine the optimal power levels across the relays and
the linear receivers are derived. Joint adaptive and iterative
recursive least squares (RLS) algorithms are also developed for
solving the optimization problems, mitigating the effects of MAI
and ISI, and allocating the power levels across the links.

This paper is organized as follows. Section II describes a
cooperative DS-CDMA system model with multiple relays. Section III
formulates the problem and the constrained linear MMSE design of
the receiver and the power allocation. The proposed RLS algorithms
for the estimation of the receive filter, the power allocation and
the channels subject to a global and individual power constraints
are developed in Sections IV and V, respectively. Section VI
presents and discusses the simulations and Section VII draws the
conclusions. \vspace{-0.15em }

\section{Cooperative DS-CDMA System Model}

Consider the uplink of a synchronous DS-CDMA system communicating
over multipath channels with QPSK modulation, $K$ users, $N$ chips
per symbol and $L$ as the maximum number of propagation paths for
each link. The network is equipped with an AF protocol that allows
communication in multiple hops using $n_r$ relays in a repetitive
fashion. We assume that the source node or terminal transmits data
organized in packets with $P$ symbols, there is enough training
and control data to coordinate transmissions and cooperation, and
the linear receivers at the relay and destination terminals are
perfectly synchronized. The received signals are filtered by a
matched filter, sampled at chip rate and organized into $M \times
1$ vectors ${\boldsymbol r}_{sd}[i]$ and ${\boldsymbol
r}_{sr_i}[i]$ which describe the signal received from the source
to the destination and from the source to the relays,
respectively, as follows
\begin{equation}
\begin{split}
{\boldsymbol r}_{sd}[n] & = \sum_{k=1}^K  a_{sd}^k[n] {\boldsymbol
D}_k {\boldsymbol h}_{sd,k}[n]b_k[i]  \\ & \quad + {\boldsymbol
\eta}_{bd}[n] +
{\boldsymbol n}_{sd}[n], \\
{\boldsymbol r}_{sr_j}[m] & = \sum_{k=1}^K a_{br_j}^k[m]
{\boldsymbol D}_k {\boldsymbol h}_{sr_j,k}[m] {b}_k[i]  \\ & \quad
+ {\boldsymbol \eta}_{sr_j}[m] +
{\boldsymbol n}_{sr_j}[m], \\
& r_j = 1,2, \ldots, n_r,~~ i=0, 1, \ldots, P-1 \\
& n=n_ri+1, ~~ m = n_ri+r_j+1 \label{rvec}
\end{split}
\end{equation}
where $M=N+L-1$, ${\boldsymbol n}_{sd}[i]$ and ${\boldsymbol
n}_{sr_j}[i]$ are zero mean complex Gaussian vectors with variance
$\sigma^2$ generated at the receiver of the destination and the
relays, and the vectors ${\boldsymbol \eta}_{sd}[i]$ and
${\boldsymbol \eta}_{sr_j}[i]$ represent the intersymbol
interference (ISI). The $M \times L$ matrix ${\boldsymbol D}_k$
has the signature sequences of each user shifted down by one
position at each column to form
\begin{equation}
{\boldsymbol D}_k = \left[\begin{array}{c c c }
d_{k}(1) &  & {\bf 0} \\
\vdots & \ddots & d_{k}(1)  \\
d_{k}(N) &  & \vdots \\
{\bf 0} & \ddots & d_{k}(N)  \\
 \end{array}\right],
\end{equation}
where ${\boldsymbol d}_k = \big[d_{k}(1), ~d_{k}(2),~ \ldots,~
d_{k}(N) \big]$ stands for the signature sequence of user $k$, the
$L \times 1$ channel vectors  from source to destination, source
to relay, and relay to destination are ${\boldsymbol
h}_{sd,k}[n]$, ${\boldsymbol h}_{r_jd,k}[n]$, ${\boldsymbol
h}_{r_js,k}[n]$, respectively. By collecting the data vectors in
(\ref{rvec}) (including the links from relays to destination) into
a $(n_r+1)M \times 1$ received vector at the destination we get {
\begin{equation*}
\begin{split}
 \left[\begin{array}{c}
  {\boldsymbol r}_{sd}[n] \\
  {\boldsymbol r}_{r_{1}d}[m] \\
  \vdots \\
  {\boldsymbol r}_{r_{n_r}d}[m]
\end{array}\right] & = \left[\begin{array}{c}
  \sum_{k=1}^K  a_{sd}^k[n] {\boldsymbol D}_k {\boldsymbol h}_{sd,k}[n]b_k[i] \\
  \sum_{k=1}^K  a_{{r_1}d}^k[m] {\boldsymbol D}_k {\boldsymbol h}_{{r_1}d,k}[m]{\tilde b}_k^{{r_1}d}[i] \\
  \vdots \\
  \sum_{k=1}^K  a_{{r_{n_r}}d}^k[m] {\boldsymbol D}_k {\boldsymbol h}_{r_{n_r}d,k}[m]{\tilde b}_k^{{r_{n_r}}d}[i]
\end{array}\right] \\ & \quad + {\boldsymbol \eta}[i] + {\boldsymbol n}[i]
\end{split}
\end{equation*}}
Rewriting the above signals in a compact form yields
\begin{equation}
{\boldsymbol r}[i] = \sum_{k=1}^{K} {\boldsymbol {\mathcal C}}_k
{\boldsymbol {\mathcal H}}_k[i] {\boldsymbol B}_k[i] {\boldsymbol
a}_k[i] + {\boldsymbol \eta}[i] + {\boldsymbol n}[i],
\label{recdata}
\end{equation}
where the $(n_r+1)M \times (n_r+1)L$ block diagonal matrix
${\boldsymbol {\mathcal C}}_k$ contains shifted versions of
${\boldsymbol D}_k$ as shown by
\begin{equation}
{\boldsymbol {\mathcal C}}_k = \left[\begin{array}{c c c c}
{\boldsymbol D}_{k} & {\bf 0} & \ldots & {\bf 0} \\
{\bf 0} & {\boldsymbol D}_{k} & \ldots & \vdots  \\
\vdots & \vdots & \ddots & {\bf 0} \\
{\bf 0} & {\bf 0} & \ldots &  {\boldsymbol D}_{k}  \\
 \end{array}\right].
\end{equation}
The $(n_r+1)L \times (n_r+1)$ matrix ${\boldsymbol {\mathcal
H}}_k[i]$ contains the channel gains of the links between the
source and the destination, and the relays and the destination.
The $(n_r+1) \times (n_r+1)$ diagonal matrix ${\boldsymbol B}_k[i]
= {\rm diag}(b_k[i]~ {\tilde b}_k^{{r_1}d}[i] \ldots {\tilde
b}_k^{{r_n}d}[i]) $ contains the symbols transmitted from the
source to the destination ($b_k[i]$) and the $n_r$ symbols
transmitted from the relays to the destination (${\tilde
b}_k^{{r_1}d}[i] \ldots {\tilde b}_k^{{r_n}d}[i]$) on the main
diagonal, the $(n_r+1) \times 1$ vector ${\boldsymbol
a}_k[i]=[a_{sd}^k[n]~a_{{r_1}d}^k[m]\ldots a_{{r_{n_r}}d}^k[m]]^T$
of the amplitudes of the links, the $(n_r+1)M \times 1$ vector
${\boldsymbol \eta}[i]$ with the ISI terms and $(n_r+1)M \times 1$
vector ${\boldsymbol n}[i]$ with the noise.

\section{Problem Formulation and Proposed MMSE Design }

This section formulates the problem of MMSE joint power allocation
and interference suppression for a cooperative DS-CDMA network.
Two constrained optimization problems are proposed in order to
describe the joint power allocation and interference suppression
problems subject to a global and individual power constraints.

\subsection{MMSE Design with Global Power Constraint}

The MMSE design of the power allocation of the links across the
source, relay and destination terminals and interference
suppression filters is presented using a global power constraint.
Let us express the received vector in (\ref{recdata}) in a more
convenient way for the proposed optimization. The $(n_r+1)M \times
1$ received vector can be written as
\begin{equation}
{\boldsymbol r}[i] = {\boldsymbol {\mathcal C}}_T {\boldsymbol
{\mathcal H}}_T[i] {\boldsymbol B}_T[i] {\boldsymbol a}_T[i] +
{\boldsymbol \eta}[i] + {\boldsymbol n}[i], \label{recdatat}
\end{equation}
where the $(n_r+1)M \times K(n_r+1)L$ matrix ${\boldsymbol
{\mathcal C}}_T = [ {\boldsymbol {\mathcal C}}_1 ~ {\boldsymbol
{\mathcal C}}_2 ~ \ldots ~  {\boldsymbol {\mathcal C}}_K ]$
contains all the signatures, the $K(n_r+1)L \times K(n_r+1)$ block
diagonal matrix ${\boldsymbol {\mathcal H}}_T[i]$ contains the
channel gains of the all links, the $K(n_r+1) \times K(n_r+1)$
diagonal matrix ${\boldsymbol B}_k[i] = {\rm diag}(b_1[i]~ {\tilde
b}_1^{{r_1}d}[i]$ \\ $\ldots {\tilde b}_1^{{r_n}d}[i]~ \ldots ~
b_K[i]~ {\tilde b}_K^{{r_1}d}[i] \ldots {\tilde b}_K^{{r_n}d}[i])$
contains the symbols transmitted from all the sources to the
destination and from all the relays to the destination on the main
diagonal, the $K(n_r+1) \times 1$ vector ${\boldsymbol
a}_T[i]=[a_{sd}^1[n]~a_{{r_1}d}^1[m]\ldots a_{{r_{n_r}}d}^1[m],~
\ldots, ~ a_{sd}^K[n]~a_{{r_1}d}^K[m]$ \\ $\ldots
a_{{r_{n_r}}d}^K[m]]^T$ of the amplitudes of all the links.

Consider an MMSE design of the receivers for the users represented
by a $(n_r+1)M \times K$ parameter matrix ${\boldsymbol W}[i]=[
{\boldsymbol w}_1[i],~ \ldots, ~ {\boldsymbol w}_K[i]]$ and for
the computation of the $K(n_r +1) \times 1$ optimal power
allocation vector ${\boldsymbol a}_{T,{\rm opt}}[i]$. This problem
is cast as
\begin{equation}
\begin{split}
[ {\boldsymbol W}_{{\rm opt}}, {\boldsymbol a}_{T,{\rm opt}}  ] &
= \arg \min_{{\boldsymbol W}[i], {\boldsymbol a}_k[i]} ~
E[ ||{\boldsymbol b}[i] - {\boldsymbol W}^H[i]{\boldsymbol r}[i] ||^2 ] \\
{\rm subject ~to~} & {\boldsymbol a}_T^H[i] {\boldsymbol a}_T[i] =
P_{T} , \label{probt}
\end{split}
\end{equation}
where the $K \times 1$ vector ${\boldsymbol b}[i] = [b_1[i], ~
\ldots, b_K[i]]^T$ represents the desired symbols. The MMSE
expressions for the parameter matrix ${\boldsymbol W}_{\rm opt}$
and vector ${\boldsymbol a}_{T, {\rm opt}}$ can be obtained by
transforming the above constrained optimization problem into an
unconstrained one with the method of Lagrange multipliers
\cite{haykin} which leads to
\begin{equation}
\begin{split}
{\mathcal L} & = E\big[ \big\|{\boldsymbol b}[i] - {\boldsymbol
W}^H[i]\big({\boldsymbol {\mathcal C}}_T {\boldsymbol {\mathcal
H}}_T[i] {\boldsymbol B}_T[i] {\boldsymbol a}_T[i] +
{\boldsymbol \eta}[i] + {\boldsymbol n}[i]\big) \big\|^2 \big] \\
& \quad + \lambda ({\boldsymbol a}_T^H[i] {\boldsymbol a}_T[i]
-P_{T}), \label{lagt}
\end{split}
\end{equation}
Fixing ${\boldsymbol a}_T[i]$, taking the gradient terms of the
Lagrangian and equating them to zero yields
\begin{equation}
{\boldsymbol W}_{{\rm opt}} = {\boldsymbol R}^{-1} {\boldsymbol
P}_{{\boldsymbol{\mathcal C}}{\boldsymbol{\mathcal H}}},
\label{wvect}
\end{equation}
where the covariance matrix of the received vector is given by
${\boldsymbol R} = E[{\boldsymbol r}[i]{\boldsymbol r}^H[i]] =
{\boldsymbol {\mathcal C}}_T {\boldsymbol {\mathcal
H}}_T[i]{\boldsymbol B}_T[i] {\boldsymbol a}_T[i] {\boldsymbol
a}_T^H[i] {\boldsymbol B}_T^H[i] {\boldsymbol {\mathcal H}}^H_T[i]
{\boldsymbol {\mathcal C}}_T^H + \sigma^2 {\boldsymbol I}$ and
${\boldsymbol P}_{{\boldsymbol{\mathcal C}}{\boldsymbol{\mathcal
H}}} = E[ {\boldsymbol r}[i]{\boldsymbol b}^H[i]] = E[{\boldsymbol
{\mathcal C}}_T {\boldsymbol {\mathcal H}}_T[i]{\boldsymbol
B}_T[i]{\boldsymbol a}_T[i] {\boldsymbol b}^H[i]] $ is the
$(n_r+1)M \times K$ cross-correlation matrix. The matrices
${\boldsymbol R}$ and ${\boldsymbol P}_{{\boldsymbol{\mathcal
C}}{\boldsymbol{\mathcal H}}}$ depend on the power allocation
vector ${\boldsymbol a}_T[i]$. The expression for ${\boldsymbol
a}_T[i]$ is obtained by fixing ${\boldsymbol W}[i]$, taking the
gradient terms of the Lagrangian and equating them to zero which
leads to
\begin{equation}
{\boldsymbol a}_{T,{\rm opt}} = ( {\boldsymbol R}_{{\boldsymbol
a}_T} + \lambda_T {\boldsymbol I})^{-1} {\boldsymbol
p}_{{\boldsymbol a}_T} \label{avect}
\end{equation}
where the $K(n_r+1) \times K(n_r+1)$ covariance matrix
${\boldsymbol R}_{{\boldsymbol a}_T} = E[{\boldsymbol {\boldsymbol
B}}_T^H[i] {\boldsymbol {\mathcal H}}^H_T[i] {\boldsymbol
{\mathcal C}}_T^H {\boldsymbol W}[i] {\boldsymbol
W}^H[i]{\boldsymbol {\mathcal C}}_T {\boldsymbol {\mathcal
H}}_T[i]{\boldsymbol B}_T[i]]$  and the vector ${\boldsymbol
p}_{{\boldsymbol a}_T} = E[{\boldsymbol B}_T^H[i] {\boldsymbol
{\mathcal H}}_T[i]^H {\boldsymbol {\mathcal C}}_T^H {\boldsymbol
W}[i] {\boldsymbol b}[i]]$ is a $K(n_r+1) \times 1$
cross-correlation vector. The Lagrange multiplier $\lambda_T$ in
the expression above plays the role of a regularization term and
has to be determined numerically due to the difficulty of
evaluating its expression. The expressions in (\ref{wvect}) and
(\ref{avect}) depend on each other and require the estimation of
the channel matrix ${\boldsymbol {\mathcal H}}_T[i]$. Thus, it is
necessary to iterate (\ref{wvect}) and (\ref{avect}) with initial
values to obtain a solution and to estimate the channel. In
addition, the network has to convey all the information necessary
to compute the global power allocation including the filter
${\boldsymbol W}_{{\rm opt}}$. The expressions in (\ref{wvect})
and (\ref{avect}) require matrix inversions with cubic complexity
( $O(((n_r+1)M)^3)$ and $O((K(n_r+1))^3)$, should be iterated as
they depend on each other and require channel estimation.

\subsection{ MMSE Design with Individual Power Constraints}

Here, it is posed an optimization problem that considers the joint
design of a linear receiver and the calculation of the optimal
power levels across the relays subject to individual power
constraints. Consider an MMSE approach for the design of the
receive filter ${\boldsymbol w}_k[i]$ and vector ${\boldsymbol
a}_k[i]$ for user $k$. This problem can be cast as
\begin{equation}
\begin{split}
[ {\boldsymbol w}_{k,{\rm opt}}, {\boldsymbol a}_{k,{\rm opt}}  ]
& = \arg \min_{{\boldsymbol w}_k[i], {\boldsymbol a}_k[i]} ~
E[ |b_k[i] - {\boldsymbol w}_k^H[i]{\boldsymbol r}[i] |^2 ] \\
{\rm subject ~to~} & {\boldsymbol a}_k^H[i] {\boldsymbol a}_k[i] =
P_{A,k} , ~~~  k   = 1,~ 2,~\ldots, ~K. \label{prob}
\end{split}
\end{equation}
The expressions for the parameter vectors ${\boldsymbol w}_k[i]$
and ${\boldsymbol a}_k[i]$ can be obtained by transforming the
above constrained optimization problem into an unconstrained one
with the help of the method of Lagrange multipliers \cite{haykin}
which leads to
\begin{equation}
\begin{split}
{\mathcal L} & = E\big[ \big|b_k[i] - {\boldsymbol
w}_k^H[i]\big(\sum_{l=1}^{K} {\boldsymbol {\mathcal C}}_l
{\boldsymbol {\mathcal H}}_l[i] {\boldsymbol B}_l[i] {\boldsymbol
a}_l[i] + {\boldsymbol \eta}[i] + {\boldsymbol n}[i]\big) \big|^2
\big] \\ & \quad + \lambda ({\boldsymbol a}_k^H[i] {\boldsymbol
a}_k[i] -P_{A,k}), ~~~  k   = 1,~ 2,~\ldots, ~K. \label{lag}
\end{split}
\end{equation}
Fixing ${\boldsymbol a}_k[i]$, taking the gradient terms of the
Lagrangian and equating them to zero yields
\begin{equation}
{\boldsymbol w}_{k,{\rm opt}} = {\boldsymbol R}^{-1} {\boldsymbol
p}_{{\boldsymbol{\mathcal C}}{\boldsymbol{\mathcal H}}},~~~  k =
1,~ 2,~\ldots, ~K, \label{wvec}
\end{equation}
where ${\boldsymbol R} = \sum_{k=1}^{K}{\boldsymbol {\mathcal
C}}_k {\boldsymbol {\mathcal H}}_k[i]{\boldsymbol B}_k[i]
{\boldsymbol a}_k[i] {\boldsymbol a}_k^H[i] {\boldsymbol B}_k^H[i]
{\boldsymbol {\mathcal H}}^H_k[i] {\boldsymbol {\mathcal C}}_k^H +
\sigma^2 {\boldsymbol I}$ is the covariance matrix and
${\boldsymbol p}_{{\boldsymbol{\mathcal C}}{\boldsymbol{\mathcal
H}}} = E[b_k^*[i] {\boldsymbol r}[i]] = {\boldsymbol {\mathcal
C}}_k {\boldsymbol {\mathcal H}}[i]{\boldsymbol a}_k[i] $ is the
cross-correlation vector. The quantities ${\boldsymbol R}$ and
${\boldsymbol p}_{{\boldsymbol{\mathcal C}}{\boldsymbol{\mathcal
H}}}$ depend on ${\boldsymbol a}_k[i]$. By fixing ${\boldsymbol
w}_k[i]$, the expression for ${\boldsymbol a}_k[i]$ is given by
\begin{equation}
{\boldsymbol a}_{k,{\rm opt}} = ( {\boldsymbol R}_{{\boldsymbol
a}_k} + \lambda {\boldsymbol I})^{-1} {\boldsymbol
p}_{{\boldsymbol a}_k}, ~~~  k   = 1,~ 2,~\ldots, ~K, \label{avec}
\end{equation}
where ${\boldsymbol R}_{{\boldsymbol a}_k} = \sum_{k=1}^{K}
{\boldsymbol {\boldsymbol B}}_k^H[i] {\boldsymbol {\mathcal
H}}^H_k[i] {\boldsymbol {\mathcal C}}_k^H {\boldsymbol w}_{k}[i]
{\boldsymbol w}_{k}^H[i]{\boldsymbol {\mathcal C}}_k {\boldsymbol
{\mathcal H}}_k[i]{\boldsymbol B}_k[i]$ is the $(n_r+1) \times
(n_r+1)$ covariance matrix and the $(n_r+1) \times 1$
cross-correlation vector is ${\boldsymbol p}_{{\boldsymbol a}_k} =
E[b_k[i] {\boldsymbol B}_k^H[i] {\boldsymbol {\mathcal H}}_k[i]^H
{\boldsymbol {\mathcal C}}_k^H
{\boldsymbol w}_k[i] ]$. 
The expressions in (\ref{wvec}) and (\ref{avec}) depend on each
other and should be iterated. They also require the estimation of
the channel matrices ${\boldsymbol {\mathcal H}}_k[i]$ and need
matrix inversions with cubic complexity ( $O(((n_r+1)M)^3)$ and
$O((n_r+1)^3)$. In what follows, we will develop algorithms for
computing ${\boldsymbol a}_{k,{\rm opt}}$, ${\boldsymbol
w}_{k,{\rm opt}}$ and the channels ${\boldsymbol {\mathcal
H}}_k[i]$ for $k=1, \ldots, K$.

\section{Proposed Joint Estimation Algorithms with a Global Power Constraint}

Here we present adaptive joint estimation algorithms to determine
the parameters of the linear receiver, the power allocation and
the channel with a global power constraint.

\subsection{Receiver and Power Allocation Parameter Estimation}

Let us now consider the following proposed least squares (LS)
optimization problem
\begin{equation}
\begin{split}
[ \hat{\boldsymbol W}[i], \hat{\boldsymbol a}_{T} [i] ] & = \arg
\min_{{\boldsymbol W}[i], {\boldsymbol a}_T[i]} ~
\sum_{l=1}^{i} \alpha^{i-l} ||{\boldsymbol b}[l] - {\boldsymbol W}^H[i]{\boldsymbol r}[l] ||^2  \\
{\rm subject ~to~} & {\boldsymbol a}_T^H[i] {\boldsymbol a}_T[i] =
P_{T},  \label{probtlst}
\end{split}
\end{equation}
where $\alpha$ is a forgetting factor. The goal is to develop a
recursive cost-effective solution to (\ref{probtlst}). To this
end, we will resort to the theory of adaptive algorithms
\cite{haykin} and derive a constrained joint iterative recursive
least squares (RLS) algorithm. This algorithm will compute
$\hat{\boldsymbol W}[i]$ and $\hat{\boldsymbol a}_T[i]$ and will
exchange information between the recursions. The part of the
algorithm to compute $\hat{\boldsymbol W}[i]$ uses ${\boldsymbol
\Phi}[i] = \hat{\boldsymbol R}[i]= \sum_{l=1}^i \alpha^{l=i}
{\boldsymbol r}[l]{\boldsymbol r}^H[l]$ and is given by
\begin{equation}
{\boldsymbol k}[i] = \frac{\alpha^{-1} {\boldsymbol \Phi}[i]
{\boldsymbol r}[i]}{1+ \alpha^{-1} {\boldsymbol r}^H[i]
{\boldsymbol \Phi}[i] {\boldsymbol r}[i]}, \label{mil1at}
\end{equation}
\begin{equation}
{\boldsymbol \Phi}[i] = \alpha^{-1} {\boldsymbol \Phi}[i-1] -
\alpha^{-1} {\boldsymbol k}[i] {\boldsymbol r}^H[i] {\boldsymbol
\Phi}[i-1], \label{mil1bt}
\end{equation}
\begin{equation} \hat{\boldsymbol W}[i] = \hat{\boldsymbol
W}[i-1] + {\boldsymbol k}[i] {\boldsymbol \xi}^H[i], \label{wrlst}
\end{equation}
where the \textit{a priori} estimation error is given by
\begin{equation}
{\boldsymbol \xi}[i] = {\boldsymbol b}[i] - \hat{\boldsymbol
W}^H[i-1] {\boldsymbol r}[i]. \label{ape1t}
\end{equation}
The derivation for the recursion that estimates the power
allocation $\hat{\boldsymbol a}_T[i]$ presents a difficulty
related to the enforcement of the constraint and how to
incorporate it into an efficient RLS algorithm. A modification is
required in order to complete the derivation of the proposed RLS
algorithm. This is because the problem in (\ref{probtlst})
incorporates a Lagrange multiplier ($\lambda$) to ensure the
global power constraint, which is difficult to include in the
matrix inversion lemma.

Our approach is to obtain a recursive expression by relaxing the
constraint and then ensure the constraint is incorporated via a
subsequent normalization procedure. In order to develop the
recursions for $\hat{\boldsymbol a}_T[i]$, we need to compute the
inverse of {\small $\hat{\boldsymbol
R}_{\boldsymbol{a}_T}[i]=\sum_{l=1}^{i}{\boldsymbol {\boldsymbol
B}}_T^H[l] \hat{\boldsymbol {\mathcal H}}^H_T[l] {\boldsymbol
{\mathcal C}}_T^H \hat{\boldsymbol W}[i] \hat{\boldsymbol
W}^H[i]{\boldsymbol {\mathcal C}}_T \hat{\boldsymbol {\mathcal
H}}_T[l]{\boldsymbol B}_T[l]$}. To this end, let us first define
${\boldsymbol \Phi}_{\boldsymbol {a}_T} = \hat{\boldsymbol
R}_{\boldsymbol{a}_T}[i]$ and employ the matrix inversion lemma
\cite{haykin} as follows:{\small
\begin{equation}
{\boldsymbol k}_{\boldsymbol{a}_T}[i] = \frac{\alpha^{-1}
{\boldsymbol \Phi}_{\boldsymbol {a}_T}[i] {\boldsymbol
B}_T^H[i]\hat{\boldsymbol{\mathcal H}}^H_T[i]{\boldsymbol
{\mathcal C}}_T^H \hat{\boldsymbol W}[i]}{1+ \alpha^{-1}
\hat{\boldsymbol W}^H[i]{\boldsymbol {\mathcal C}}_T
\hat{\boldsymbol{\mathcal H}}_T[i]{\boldsymbol B}_T[i]
{\boldsymbol \Phi}_{\boldsymbol {a}_T}[i] {\boldsymbol
B}_T^H[i]\hat{\boldsymbol{\mathcal H}}^H_T[i]{\boldsymbol
{\mathcal C}}_T^H \hat{\boldsymbol W}[i]} \label{mil2at}
\end{equation}}
{\small
\begin{equation}
{\boldsymbol \Phi}_{\boldsymbol {a}_T}[i] = \alpha^{-1}
{\boldsymbol \Phi}_{\boldsymbol {a}_T}[i-1] - \alpha^{-1}
{\boldsymbol k}_{\boldsymbol{a}_T}[i] \hat{\boldsymbol
W}^H[i]{\boldsymbol {\mathcal C}}_T \hat{\boldsymbol{\mathcal
H}}_T[i]{\boldsymbol B}_T[i] {\boldsymbol \Phi}_{\boldsymbol
{a}_T}[i-1] \label{mil2bt}
\end{equation}}
After some algebraic manipulations, we get
\begin{equation}
\hat{\boldsymbol a}_T[i] = \hat{\boldsymbol a}_T[i] + {\boldsymbol
k}_{\boldsymbol{a}_T}[i] \xi_{\boldsymbol{a}_T}^*[i]
\end{equation}
where the \textit{a priori} estimation error for this recursion is
\begin{equation}
\xi_{\boldsymbol{a}_T}[i] = {\boldsymbol b}[i] - \hat{\boldsymbol
a}_T^H[i]{\boldsymbol B}_k^H[i]\hat{\boldsymbol{\mathcal
H}}^H_T[i]{\boldsymbol {\mathcal C}}_T^H \hat{\boldsymbol W}[i]
\end{equation}
In order to ensure the individual power constraint ${\boldsymbol
a}_T^H[i]{\boldsymbol a}_T[i] = P_{T}$, we apply the following
rule
\begin{equation}
\hat{\boldsymbol a}_T[i] \leftarrow P_{T} ~ \hat{\boldsymbol
a}_T[i] \big(\sqrt{\hat{\boldsymbol a}_T^H[i]\hat{\boldsymbol
a}_T[i]}\big)^{-1}
\end{equation}
The algorithms for recursive computation of $\hat{\boldsymbol
W}[i]$ and ${\boldsymbol a}_T[i]$ require estimates of the channel
vector ${\boldsymbol {\mathcal H}}_T[i]$, which will also be
developed in what follows. The complexity of the proposed
algorithm is $O(((n_r+1)M)^2)$ for calculating $\hat{\boldsymbol
W}[i]$ and $O((K(n_r+1))^2)$ for obtaining $\hat{\boldsymbol
a}_T[i]$.

\subsection{Channel Estimation with a Global Power Constraint}

We present a channel estimator that considers jointly all the $K$
users and exploits the knowledge of the receive filter matrix
$\hat{\boldsymbol W}[i]$ and the power allocation vector
$\hat{\boldsymbol a}_T[i]$. Let us rewrite ${\boldsymbol r}[i]$ as
\begin{equation}
{\boldsymbol r}[i] = {\boldsymbol {\mathcal C}}_T
{\breve{\boldsymbol B}}_T[i] {\breve{\boldsymbol A}}_T[i]
{\boldsymbol h}_T[i] + {\boldsymbol \eta}[i] + {\boldsymbol n}[i],
\label{recdatah}
\end{equation}
where ${\breve{\boldsymbol B}}_T[i]$ is a $K(n_r+1)L \times
K(n_r+1)L$ matrix with the symbols of all $K$ users transmitted
from the sources and the relays on the main diagonal,
${\breve{\boldsymbol A}}_T[i]$ is a $K(n_r+1)L \times K(n_r+1)L$
matrix with the amplitudes of all the links on the main diagonal
and ${\boldsymbol h}_T[i]$ is a $K(n_r+1)L \times 1$ vector with
the $L$ channel gains from all $K$ users, sources and relays. A
channel estimator can be derived from the following optimization
problem
\begin{equation}
\begin{split}
\hat{\boldsymbol h}_T[i] & = \arg \min_{{\boldsymbol h}_T[i]} ~
\sum_{l=1}^{i} \alpha^{i-l} || {\boldsymbol r}[l] - {\boldsymbol
{\mathcal C}}_T {\breve{\boldsymbol B}}_T[i] {\breve{\boldsymbol
A}}_T[i] {\boldsymbol h}_T[i] ||^2
\end{split}
\end{equation}
The solution to the above optimization problem is given by
\begin{equation}
\hat{\boldsymbol h}_T[i] = \hat{\boldsymbol R}_{{\boldsymbol
h}_T}^{-1}[i] \hat{\boldsymbol p}_{{\boldsymbol h}_T}[i],
\label{cestt}
\end{equation}
where
\begin{equation}
\hat{\boldsymbol R}_{{\boldsymbol h}_T}^{-1}[i] = \alpha^{-1}
\hat{\boldsymbol R}_{{\boldsymbol h}_T}^{-1}[i-1] -
\frac{\alpha^{-2} \hat{\boldsymbol R}_{{\boldsymbol
h}_T}^{-1}[i-1] {\boldsymbol v}_T[i] {\boldsymbol v}_T^H[i]
\hat{\boldsymbol R}_{{\boldsymbol h}_T}^{-1}[i-1]}{1 + \alpha^{-1}
{\boldsymbol v}_T^H[i] \hat{\boldsymbol R}_{{\boldsymbol
h}_T}^{-1}[i-1] {\boldsymbol v}_T[i]}, \label{milrt}
\end{equation}
\begin{equation}
\hat{\boldsymbol p}_{{\boldsymbol h}_T}[i] = \alpha
\hat{\boldsymbol p}_{{\boldsymbol h}_T}[i-1] + {\boldsymbol
v}_T^H[i] {\boldsymbol r}[i]
\end{equation}
and
\begin{equation}
{\boldsymbol v}_T[i] = {\boldsymbol {\mathcal C}}_T
{\breve{\boldsymbol B}}_T[i] {\breve{\boldsymbol A}}_T[i]
\end{equation}
The channel matrix $\hat{\boldsymbol {\mathcal H}}_T[i]$ is
determined with the recursion:
\begin{equation}
\hat{\boldsymbol {\mathcal H}}_T[i] = \sum_{j=1}^{K(n_r+1)}
\hat{\boldsymbol h}_T[i] {\boldsymbol q}_{T,j},
\end{equation}
where ${\boldsymbol q}_{T,j} =
[\underbrace{0~\ldots~0}_{j-1}~1~\underbrace{0~\ldots~0}_{K(n_r+1)-j}]$.

This algorithm jointly estimates the coefficients of the channels
across all the links and for all users subject to a global power
constraint. The complexity of the proposed RLS channel estimation
algorithm is $O((K(n_r+1)L)^2)$.

\section{Proposed Joint Estimation Algorithms with Individual Power
Constraints}

In this section, we present adaptive algorithms to determine the
parameters of the linear receiver, the power allocation and the
channel subject to individual power constraints. Unlike the
algorithms presented in the previous section, the algorithms
proposed here are suitable for distributed resource allocation,
detection and estimation.

\subsection{Receiver and Power Allocation Parameter
Estimation}

We develop a recursive solution to the expressions in (\ref{wvec})
and (\ref{avec}) using time averages instead of the expected
value. The proposed RLS algorithms will compute $\hat{\boldsymbol
w}_k[i]$ and $\hat{\boldsymbol a}_k[i]$ for each user $k$ and will
exchange information between the recursions. We fix
$\hat{\boldsymbol a}_k[i]$ and compute the inverse of
$\hat{\boldsymbol R}[i]$ using the matrix inversion lemma
\cite{haykin} to obtain $\hat{\boldsymbol w}_k[i]$. Defining
${\boldsymbol \Phi}[i] = \hat{\boldsymbol R}[i]$ then we can
obtain the recursions
\begin{equation}
{\boldsymbol k}[i] = \frac{\alpha^{-1} {\boldsymbol \Phi}[i]
{\boldsymbol r}[i]}{1+ \alpha^{-1} {\boldsymbol r}^H[i]
{\boldsymbol \Phi}[i] {\boldsymbol r}[i]} \label{mil1a}
\end{equation}
\begin{equation}
{\boldsymbol \Phi}[i] = \alpha^{-1} {\boldsymbol \Phi}[i-1] -
\alpha^{-1} {\boldsymbol k}[i] {\boldsymbol r}^H[i] {\boldsymbol
\Phi}[i-1] \label{mil1b}
\end{equation}
\begin{equation}
{\boldsymbol w}_k[i] = {\boldsymbol w}_k[i-1] + {\boldsymbol k}[i]
\xi^*[i] \label{wrls}
\end{equation}
where the \textit{a priori} estimation error is
\begin{equation}
\xi[i] = b_k[i] - {\boldsymbol w}_k^H[i-1] {\boldsymbol r}[i].
\label{ape1}
\end{equation}
The derivation for the recursion that estimates the power
allocation follows a similar approach to the computation of
$\hat{\boldsymbol w}_k[i]$. In order to develop the recursions for
$\hat{\boldsymbol a}_k[i]$, we need to compute the inverse of
$\hat{\boldsymbol R}_{\boldsymbol{a}_k}[i] =\sum_{l=1}^{i}
{\boldsymbol {\boldsymbol B}}_k^H[l] \hat{\boldsymbol {\mathcal
H}}^H_k[l] {\boldsymbol {\mathcal C}}_k^H \hat{\boldsymbol w}_k[l]
\hat{\boldsymbol w}^H_k[l]{\boldsymbol {\mathcal C}}_k
\hat{\boldsymbol {\mathcal H}}_k[l]{\boldsymbol B}_k[l]$. To this
end, let us first define ${\boldsymbol \Phi}_{\boldsymbol {a}_k} =
\hat{\boldsymbol R}_{\boldsymbol{a}_k}[i]$ and proceed as follows:
\begin{equation}
{\boldsymbol k}_{\boldsymbol{a}_k}[i] = \frac{\alpha^{-1}
{\boldsymbol \Phi}_{\boldsymbol {a}_k}[i] {\boldsymbol
B}_k^H[i]\hat{\boldsymbol{\mathcal H}}^H_k[i]{\boldsymbol
{\mathcal C}}_k^H \hat{\boldsymbol w}_k[i]}{1+ \alpha^{-1}
\hat{\boldsymbol w}_k^H[i]{\boldsymbol {\mathcal C}}_k
\hat{\boldsymbol{\mathcal H}}_k[i]{\boldsymbol B}_k[i]
{\boldsymbol \Phi}_{\boldsymbol {a}_k}[i] {\boldsymbol
B}_k^H[i]\hat{\boldsymbol{\mathcal H}}^H_k[i]{\boldsymbol
{\mathcal C}}_k^H \hat{\boldsymbol w}_k[i]} \label{mil2a}
\end{equation}
\begin{equation}
{\boldsymbol \Phi}_{\boldsymbol {a}_k}[i] = \alpha^{-1}
{\boldsymbol \Phi}_{\boldsymbol {a}_k}[i-1] - \alpha^{-1}
{\boldsymbol k}_{\boldsymbol{a}_k}[i] \hat{\boldsymbol
w}_k^H[i]{\boldsymbol {\mathcal C}}_k \hat{\boldsymbol{\mathcal
H}}_k[i]{\boldsymbol B}_k[i] {\boldsymbol \Phi}_{\boldsymbol
{a}_k}[i-1] \label{mil2b}
\end{equation}
\begin{equation}
\hat{\boldsymbol a}_k[i] = \hat{\boldsymbol a}_k[i] + {\boldsymbol
k}_{\boldsymbol{a}_k}[i] \xi_{\boldsymbol{a}_k}^*[i]
\end{equation}
where the \textit{a priori} estimation error for this equation is
\begin{equation}
\xi_{\boldsymbol{a}_k}[i] = b_k[i] - \hat{\boldsymbol
a}_k^H[i]{\boldsymbol B}_k^H[i]\hat{\boldsymbol{\mathcal
H}}^H_k[i]{\boldsymbol {\mathcal C}}_k^H \hat{\boldsymbol w}_k[i]
\end{equation}
In order to ensure the individual power constraint ${\boldsymbol
a}_k^H[i]{\boldsymbol a}_k[i] = P_{A,k}$, we apply the rule
\begin{equation}
\hat{\boldsymbol a}_k[i] \leftarrow P_{A,k} ~ \hat{\boldsymbol
a}_k[i] (\sqrt{\hat{\boldsymbol a}_k^H[i]\hat{\boldsymbol
a}_k[i]})^{-1}
\end{equation}
The algorithms for recursive computation of
$\hat{\boldsymbol w}_k[i]$ and $\hat{\boldsymbol a}_k[i]$ require
estimates of the channel vector ${\boldsymbol {\mathcal H}}_k[i]$,
which will also be developed in what follows. The complexity of
the proposed algorithm is $O(((n_r+1)M)^2)$ for calculating
$\hat{\boldsymbol w}_k[i]$ and $O((n_r+1)^2)$ for obtaining
$\hat{\boldsymbol a}_k[i]$.

\subsection{Channel Estimation with Individual Power Constraints}

We propose here an algorithm that estimates the channels for each
user $k$ across all links subject to individual power constraints
and exploits the knowledge of the receiver filter
$\hat{\boldsymbol w}_k[i]$ and the power allocation vector
$\hat{\boldsymbol a}_k[i]$. Let us express the received vector in
(3) as
\begin{equation}
{\boldsymbol r}[i] = \sum_{k=1}^{K} {\boldsymbol {\mathcal C}}_k
{\breve{\boldsymbol B}}_k[i] {\breve{\boldsymbol A}}_k[i]
{\boldsymbol h}_k[i] + {\boldsymbol \eta}[i] + {\boldsymbol n}[i],
\label{recdatah}
\end{equation}
where ${\breve{\boldsymbol B}}_k[i]$ is a $(n_r+1)L \times
(n_r+1)L$ matrix with the symbols of user $k$ transmitted from the
sources and the relays on the main diagonal, ${\breve{\boldsymbol
A}}_k[i]$ is a $(n_r+1)L \times (n_r+1)L$ matrix with the
amplitudes of all the links on the main diagonal and ${\boldsymbol
h}_k[i]$ is a $K(n_r+1)L \times 1$ vector with the $L$ channel
gains for user $k$, and each link. A channel estimation algorithm
can be developed solving the optimization problem
\begin{equation}
\begin{split}
\hat{\boldsymbol h}_k[i] & = \arg \min_{{\boldsymbol h}_k[i]} ~
\sum_{l=1}^{i} \alpha^{i-l} || {\boldsymbol r}[l] - {\boldsymbol
{\mathcal C}}_k {\breve{\boldsymbol B}}_k[i] {\breve{\boldsymbol
A}}_k[i] {\boldsymbol h}_k[i] ||^2, \\  ~~ {\rm for}~~ k &
=1,~2,~\ldots,K
\end{split}
\end{equation}
The solution to the above optimization problem is
\begin{equation}
\hat{\boldsymbol h}_k[i] = \hat{\boldsymbol R}_{{\boldsymbol
h}_k}^{-1}[i] \hat{\boldsymbol p}_{{\boldsymbol h}_k}[i], ~~ {\rm
for}~~ k=1,~2,~\ldots,K \label{cestk}
\end{equation}
where
\begin{equation}
\hat{\boldsymbol R}_{{\boldsymbol h}_k}^{-1}[i] = \alpha^{-1}
\hat{\boldsymbol R}_{{\boldsymbol h}_k}^{-1}[i-1] -
\frac{\alpha^{-2} \hat{\boldsymbol R}_{{\boldsymbol
h}_k}^{-1}[i-1] {\boldsymbol v}_k[i] {\boldsymbol v}_k^H[i]
\hat{\boldsymbol R}_{{\boldsymbol h}_k}^{-1}[i-1]}{1 + \alpha^{-1}
{\boldsymbol v}_k^H[i] \hat{\boldsymbol R}_{{\boldsymbol
h}_k}^{-1}[i-1] {\boldsymbol v}_k[i]}, \label{milrk}
\end{equation}
\begin{equation}
\hat{\boldsymbol p}_{{\boldsymbol h}_k}[i] = \alpha
\hat{\boldsymbol p}_{{\boldsymbol h}_k}[i-1] + {\boldsymbol
v}_k^H[i] {\boldsymbol r}[i]
\end{equation}
and
\begin{equation}
{\boldsymbol v}_k[i] = {\boldsymbol {\mathcal C}}_k
{\breve{\boldsymbol B}}_k[i] {\breve{\boldsymbol A}}_k[i]
\end{equation}
The channel matrix $\hat{\boldsymbol {\mathcal H}}_k[i]$ is
obtained as follows:
\begin{equation}
\hat{\boldsymbol {\mathcal H}}_k[i] = \sum_{j=1}^{n_r+1}
\hat{\boldsymbol h}_k[i] {\boldsymbol q}_{k,j},
\end{equation}
where ${\boldsymbol q}_{k,j} =
[\underbrace{0~\ldots~0}_{j-1}~1~\underbrace{0~\ldots~0}_{n_r+1-j}]$.

This algorithm estimates the coefficients of the channels of each
user $k$ across all the links subject to individual power
constraints. The complexity is $O(((n_r+1)L)^2)$.

\section{Simulations}

We evaluate the bit error ratio (BER) performance of the proposed
joint power allocation and interference suppression (JPAIS)
algorithms with global (GPC) and individual power (IPC)
constraints and compare them with interference suppression schemes
without cooperation (NCIS) and with cooperation (CIS) using an
equal power allocation across the relays. We consider a stationary
DS-CDMA network with randomly generated spreading codes with a
processing gain $N=16$. The fading channels are generated
considering a random power delay profile with gains taken from a
complex Gaussian variable with unit variance and mean zero, $L=3$
paths spaced by one chip, and are normalized for unit power. The
power constraint parameter $P_{A,k}$ is set for each user so that
one can control the SNR (${\rm SNR} = P_{A,k}/\sigma^2$) and
$P_T=K P_{A,k}$, whereas it follows a log-normal distribution for
with associated standard deviation of $3$ dB. We adopt the AF
cooperative strategy with repetitions and all the relays and the
destination terminal are equipped with linear MMSE receivers. We
employ packets with $1500$ QPSK symbols and average the curves
over $1000$ runs. The receivers have either full knowledge of the
channel and the noise variance or are adaptive and estimate all
the required coefficients and the channels using the proposed RLS
algorithms with optimized parameters. For the adaptive receivers,
we provide training sequences with $200$ symbols placed at the
preamble of the packets. After the training sequence, the adaptive
receivers are switched to decision-directed mode.

\begin{figure}[!htb]
\begin{center}
\def\epsfsize#1#2{0.95\columnwidth}
\epsfbox{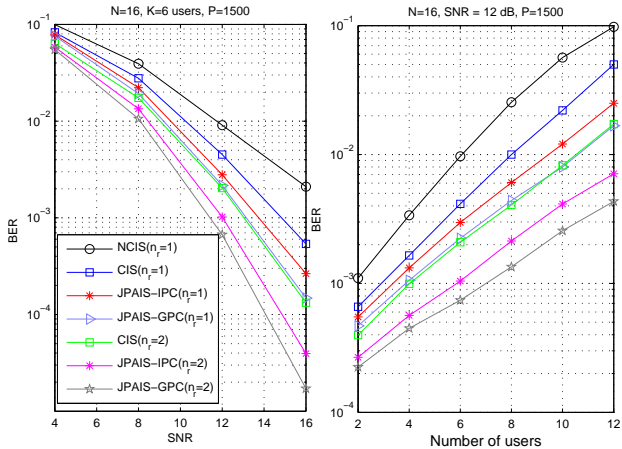} \vspace{-1.0em} \caption{\footnotesize BER versus
SNR and $K$ for the optimal linear MMSE detectors. Parameters:
$\lambda_T = \lambda =0.025$.} \vspace{-0.75em}\label{fig1}
\end{center}
\end{figure}

We first consider the proposed JPAIS method with the MMSE
expressions of (\ref{wvect}) and (\ref{avect}) using a global
power constraint (JPAIS-GPC), and (\ref{wvec}) and (\ref{avec})
with individual power constraints (JPAIS-IPC). We compare the
proposed scheme with a non-cooperative approach (NCIS) and a
cooperative scheme with equal power allocation (CIS) across the
relays for $n_r=1,2$ relays. The results shown in Fig. \ref{fig1}
illustrate the performance improvement achieved by the proposed
JPAIS scheme and algorithms, which significantly outperform the
CIS and the NCIS techniques. As the number of relays is increased
so is the performance, reflecting the exploitation of the spatial
diversity. In the scenario studied, the proposed JPAIS-IPC
approach can accommodate up to $3$ more users as compared to the
CIS scheme and double the capacity as compared with the NCIS for
the same performance. The proposed JPAIS-GPC is superior to the
JPAIS-IPC and can accommodate up to $2$ more users than the
JPAI-GPS, while its complexity is higher.

\begin{figure}[!htb]
\begin{center}
\def\epsfsize#1#2{0.95\columnwidth}
\epsfbox{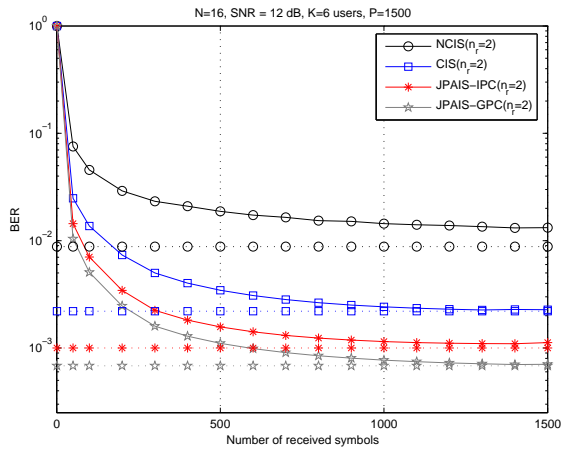} \vspace{-1.0em}\caption{\footnotesize BER
performance versus number of symbols. The curves for the adaptive
schemes are in solid lines, whereas those of the optimal MMSE
schemes are in dotted lines. Parameters: $\lambda_T=\lambda=0.025$
(for MMSE schemes), $\alpha=0.998$ (for adaptive schemes).}
\vspace{-0.75em}\label{fig3}
\end{center}
\end{figure}

The second experiment depicted in Fig. \ref{fig3} shows the BER
performance of the proposed adaptive algorithms (JPAIS) against
the existing NCIS and CIS schemes with $n_r=2$ relays. All
techniques employ RLS algorithms for the estimation of the
coefficients of the channel, the receiver filters and the power
allocation for each user (JPAIS only). The complexity of the
proposed algorithms is quadratic with the filter length of the
receivers and the number of relays $n_r$, whereas the optimal MMSE
schemes require cubic complexity. From the results, we can verify
that the proposed adaptive joint estimation algorithms converge to
approximately the same level of the MMSE schemes, which have full
channel and noise variance knowledge. Again, the proposed
JPAIS-GPC is superior to the JPAIS-IPC but requires higher
complexity and joint demodulation of signals, whereas the
JPAIS-IPC renders itself to a distributed implementation.
\vspace{-0.05em}

\section{Concluding remarks}
 \vspace{-0.05em}
This work proposed joint iterative power allocation and
interference mitigation techniques for DS-CDMA networks with
multiple relays and the AF cooperation strategy. A joint
constrained optimization framework and algorithms that consider
the allocation of power levels across the relays subject to global
and individual power constraints and the design of linear
receivers for interference suppression were proposed. The results
of simulations showed that the proposed methods obtain significant
gains in performance and capacity over existing non-cooperative
systems and cooperative schemes.

\end{document}